\documentstyle[prd,epsfig,aps,preprint]{revtex}

\begin{document}

\title {\Large Hamilton-Jacobi Approach for Power-Law Potentials}

\author {R. C. Santos$^{1,}$ \footnote{rose@dfte.ufrn.br}, J.
Santos$^{1,}$\footnote{janilo@dfte.ufrn.br}, J.  A.  S.
Lima$^{1,2,}$ \footnote{limajas@astro.iag.usp.br}}

\address{$^{1}$Departamento de F\'{\i}sica, UFRN, 59072-970, Natal, RN, Brasil
\\$^{2}$Instituto de Astronomia, Geof\'{\i}sica e Ci\^encias
Atmosf\'ericas - USP\\ Rua do Mat\~ao, 1226 - Cid.
Universit\'aria, 05508-900, S\~ao Paulo, SP, Brasil\\}
\date{\today}
\maketitle

\begin{abstract}
The classical and relativistic Hamilton-Jacobi approach is applied
to the one-dimensional homogeneous potential, $V(q)=\alpha q^n$,
where $\alpha$ and $n$ are continuously varying parameters. In the
non-relativistic case, the exact analytical solution is determined
in terms of $\alpha$, $n$ and the total energy $E$. It is also
shown that the non-linear equation of motion can be linearized by
constructing a hypergeometric differential equation for the
inverse problem $t(q)$. A variable transformation reducing the
general problem to that one of a particle subjected to a linear
force is also established. For any value of $n$, it leads to a
simple harmonic oscillator if $E>0$, an ``anti-oscillator" if
$E<0$, or a free particle if $E=0$. However, such a reduction is
not possible in the relativistic case. For a bounded relativistic
motion, the first order correction to the period is determined for
any value of $n$. For $n >> 1$, it is found that the correction is
just twice that one deduced for the simple harmonic oscillator
($n=2$), and does not depend on the specific value of $n$.
\end{abstract}

\newpage

\section{Introduction}
\setcounter{equation}{0} The Hamilton-Jacobi (HJ) equation is a
powerful method either to the relativistic and classical
framework, and also plays a prominent role in quantum mechanics as
a special route for a comprehension of the Schr\"odinger
equation\cite{goldstein,Holland}. The associated action variable
method, which is a slight modification of the HJ approach, is also
quite useful for determining frequencies and energies of periodic
systems. Even for the general relativity theory, the importance of
the HJ approach has long been recognized by many authors (see, for
instance, \cite{Carter} and Refs. therein). Indeed, one of the
most elegant methods for describing geodesics and orbits of test
particles in Schwarzchild and Kerr spacetimes, as well as for any
stationary gravitational configuration, is provided by the
relativistic HJ equation[3-5].

In the nonrelativistic domain, the standard applications of the HJ
theory are the harmonic oscillator, the Kepler problem, and
charged particles moving in electro-magnetic fields. Recently, the
method has also been applied to the rocket problem\cite{Valle96}.
Pars\cite{pars} studied the motion of a classical particle in a
plane under central attraction derived from the potential,
$V(r)=\alpha r^{n}$, for some particular values of $n$. However,
to the best of our knowledge, if $\alpha$ and $n$ are continuously
varying parameters the general solution has not been obtained even
for the one-dimensional case.

In this letter we discuss an analytical solution for the power law
potential both for the classical and relativistic case. As we
shall see, the HJ equation for a particle moving under the action
of a one-dimensional potential, $V(q)=\alpha q^{n}$, has a general
and unified solution in terms of hypergeometric functions. The
period of the motion as a function of the power index $n$ and of
the $\alpha$ parameter are easily determined from the general
solution. For completeness, we also show that such a problem may
also be exactly solved starting from the nonlinear equation of
motion by employing a slightly modified Euler-Lagrange approach.
By changing the coordinate and adopting an auxiliary time, we show
that the motion of the particle in such a potential may be reduced
to the problem of a particle under the action of a linear force:
an oscillator if $E > 0$, an ``anti-oscillator" if $E < 0$ or a
free-particle if $E = 0$, where $E$ is the total energy of the
particle. Classically, these results hold regardless of the value
of $n$. Such a reduction cannot be implemented in the relativistic
case. However, the general problem may be reduced to an integral,
which is a natural extension of the earlier Synge's treatment for
the relativistic harmonic oscillator\cite{synge}.

\section{Hamilton-Jacobi Approach}
\label{jacobi} The classical Lagrangian for a particle subject to
the one-dimensional potential, $V(q)=\alpha q^{n}$, reads
\begin{equation}\label{lagrangian}
L(q,\dot{q},t)=\frac {1}{2}\,m\dot{q}^2 - \alpha q^n \,,
\end{equation}
where an overdot means total derivative. The parameters $\alpha$,
$n$, vary continuously, and $m$, $q$ are, respectively, the mass
of the particle and its generalized coordinate. The Hamiltonian is
given by ($p$ is the canonical momentum)
\begin{equation}\label{H}
H(q,p)=\frac{p^2}{2m} + \alpha q^n=E \,,
\end{equation}
and since it does not depend explicitly on the time, it represents
a conserved quantity which is the energy of the system. The
Hamilton-Jacobi equation for the Hamiltonian (\ref{H}) assume the
following form \cite{goldstein}
\begin{equation}
\label{HJ} \frac{1}{2m}\left(\frac{\partial S}{\partial
q}\right)^2 + \alpha q^n + \frac{\partial S}{\partial t} = 0 \,,
\end{equation}
where $S$ is the Hamilton's principal function. Since the explicit
dependence of $S$ on time is involved only in the last term, the
variables can be separated. Following standard lines, the solution
is assumed to have the form
\begin{equation} \label{S} S(q,E,t) = W(q,E) - Et\,,
\end{equation}
where $E$, the constant of integration, has been identified with
the total energy. With this choice, the time is readily eliminated
from (\ref{HJ}), and Hamilton's principal function $S$ becomes

\begin{equation}\label{HPF}
 S = \int\sqrt{2mE -
2m\alpha q^n}\,dq -Et \,.
\end{equation}

It should be noticed that if the energy $E$ is positive, the
constant $\alpha$ appearing in the potential may assume either
negative or positive values, however, if $E \le 0$, the allowed
values of $\alpha$ are necessarily negative. Since $H=E$, such
considerations also follows naturally from the positivity of the
kinetic energy
\begin{equation}
\label{KE} K = \frac{1}{2}\,m\dot{q}^2 = E - \alpha q^n \,.
\end{equation}

The second integration constant is a consequence of the Jacobi
transformation equation
\begin{equation}
\label{beta-t} \beta = \frac{\partial S}{\partial E} =
\sqrt{\frac{m}{2|E|}}\int_{0}^{q} \frac{dq}{\sqrt{1 -
\frac{\alpha}{E}\,q^n}} - t\,.
\end{equation}
The root of $|E|$ arises naturally if one takes for negative
values of $E$, $E^{*}= -E$, and compute the partial derivative
with respect to $E$. In order to integrate (\ref{beta-t}) we make
the change of variable $u=(\alpha /E)q^n$. Thus we have that
\begin{equation}\label{SOLI}
\label{beta} t + \beta = \sqrt{\frac{m}{2n^2|E|}}\,\left(
\frac{E}{\alpha}\right)^{1/n} \int_{0}^{u} \frac{y^{1/n
-1}\,dy}{\sqrt{1 - y}} \:.
\end{equation}
It is worth noticing that the integral appearing on the right hand
side of the above expression is the incomplete Beta
function\cite{abramowitz}
\[
B_u(a,b)=\int_0^u\,y^{a-1}(1-y)^{b-1}dy
\]
for $a=1/n$ and $b=1/2$. Without loss of generality we may take
$q(t=0)=0$, so that $\beta =0$ in (\ref{beta}). When the motion is
finite, the particle moves back and forth (oscillatory motion),
and the period $T$ can be easily determined. We see from
(\ref{KE}) that at the turning points ($\dot{q}=0$) the amplitude
of the oscillation is $A = (E/\alpha)^{1/n}$, and, therefore,
$u_{max}=1$, is the maximum value in the upper limit of the
integral (\ref{SOLI}). The period of the motion is given by
$T=4t_{max}$ and, taking these considerations into account, we
find that equation (\ref{beta}) yields
\begin{equation}
\label{period} T= 2\sqrt{\frac{2m}{\alpha n^2}}\, A^{1-1/n}\,
B(1/n,1/2)\,,
\end{equation}
where $B(1/n,1/2) ={\sqrt\pi}\,\Gamma\left({{1 \over
n}}\right)/\Gamma\left({{1 \over n} + {1 \over 2}}\right)$ is the
complete Beta function.  The period of this oscillator is clearly
dependent on the amplitude $A$, unless $n=2$. As should be
expected, for $n=2$ equation (\ref{period}) reduces to
\[
T = 2\pi \sqrt{\frac{m}{2\alpha}}\,,
\]
which is the well known result for the simple harmonic oscillator.

It should be recalled that the incomplete Beta function is related
to the hypergeometric Gaussian $F(a,b;c;u)$ by the following
identity: $B_u(a,b)=a^{-1}u^aF(a,1-b;1+a;u)$. It thus follows that
the result (\ref{beta}) may be expressed as
\begin{equation} \label{hiper}
t + \beta = \sqrt{\frac{m}{2|E|}}\:q F\left(
\frac{1}{2},\frac{1}{n};\frac{1}{n}+
1;\frac{\alpha}{E}\,q^n\right)\;.
\end{equation}
As a check, we notice that for $n=2$ the above expression can be
written as
\begin{equation}
\label{hiper2} \omega (t + \beta)=zF\left(
\frac{1}{2},\frac{1}{2};\frac{3}{2};z^2\right)\,,
\end{equation}
where we have defined the variable $z=\sqrt{\alpha/E}\,q$ and
$\omega = 2\pi/T$ is the frequency of the oscillatory motion.
Since $F(1/2,1/2;3/2;z^2) = z^{-1}\arcsin z$ \cite{abramowitz}; by
returning to the old variable $q$, the above equation can be
recast as
\begin{equation}
\label{oscilador} q(t) =
\sqrt{\frac{2E}{m\omega^2}}\,\sin{\omega(t + \beta)}\,,
\end{equation}
which describes the motion of a simple harmonic oscillator
\cite{goldstein}.

\section{Modified Euler-Lagrange Approach}
\label{euler} We present here another method of solution for the
problem outlined in the introduction. In principle, by considering
that the HJ approach leads to the hypergeometric function for
$t(q)$ as given by (\ref{hiper}), it should be possible to obtain
the differential hypergeometric equation starting directly from
the equation of motion. The importance of this method is two-fold:
the initial conditions remains arbitrary and the linearization
procedure is a guarantee that we get the complete solution for the
nonlinear differential equation governing the motion of the
particle.

The equation of motion (Euler-Lagrange equation) for the
Lagrangian given by (\ref{lagrangian}) is
\begin{equation}
\label{eqmove} m\ddot{q} + n\alpha q^{n-1} = 0\,.
\end{equation}
Instead to solve the equation of motion above, we will consider an
equivalent equation. By multiplying it by $q$, and inserting the
definition of the total energy of the particle given by
(\ref{KE}), we obtain:
\begin{equation}
\label{eqmotion} q\ddot{q} - \frac{n}{2}\,\dot{q}^2 + \frac{nE}{m}
= 0 \,.
\end{equation}
It is worth mentioning that, if some few identifications are made,
(\ref{eqmotion}) is the same differential equation describing the
evolution of the scale factor in the standard
Friedman-Robertson-Walker (FRW) cosmological model \cite{limajas}.

A first integral of (\ref{eqmotion}) is
\begin{equation}
\label{firsti} \dot{q}^2 =
\frac{2E}{m}\left(1-\frac{\alpha}{E}\,q^n\right)\,.
\end{equation}
It should be noted that the signals for the total energy $E$ and
for the constant $\alpha$ must be properly chosen. So, if $E>0$
the constant $\alpha$ can have any signal, but if $E\leq 0$ we
must have a negative $\alpha$. The integration of (\ref{firsti})
is achieved with the introduction of an auxiliary variable defined
by $u=(\alpha /E)q^n$. With the aid of this transformation the
first integral is written as
\begin{equation}
\label{firstu} \frac{du}{dt} = nB\,u^{1-1/n}(1-u)^{1/2}\,,
\end{equation}
where
\[
B = \sqrt{\frac{2\alpha}{m}}\left(\frac{E}{\alpha}\right)^{1/2 -
1/n}\,.
\]
At this stage we could integrate (\ref{firstu}), as it was done in
Sec. (\ref{jacobi}). Instead, on may consider the inverse problem,
i.e., the solutions for a differential equation expressing the
time as a function of the auxiliary variable $u(q)$. This
inversion is readily done through (\ref{firstu}), which provides
the first derivative of $t(u)$:
\begin{equation}
\label{inverseu} \frac{dt}{du} = \frac{1}{nB}\,u^{1/n-
1}(1-u)^{-1/2}\,,
\end{equation}
whereas the second derivative may be put in the form
\begin{equation}
\label{hyperg} u(1-u)\frac{d^2t}{du^2} + \left[ c - (a +
1)u\right]\,\frac{dt}{du} = 0 \,.
\end{equation}
where $c = 1 - 1/n$ and $a=1/2 - 1/n$. This is Gauss's
hypergeometric differential equation. Its general solution
consists of a linear combination of two linearly independent
solutions \cite{seaborn}:
\begin{equation}
\label{solgeral} t(u) = C_1F(a,0;c;u) +
C_2u^{1-c}F(1+a-c,1-c;2-c;u)\,,
\end{equation}
where $F$ is the hypergeometric function, and $C_1,\;C_2$ are the
arbitrary constants of integration to be determined by the
boundary conditions. It is a property of the hypergeometric
function that if any of the first two parameters is zero, then the
series terminates (i.e., $F(a,0;c;u)=1$). It thus follows that the
solution (\ref{solgeral}) reduces to
\begin{equation}
\label{hyperF} t = C_1 +
C_2\left(\frac{\alpha}{E}\right)^{1/n}\,q\,
F\left(\frac{1}{2},\frac{1}{n};\frac{1}{n} +
1;\frac{\alpha}{E}\,q^n\right)\,,
\end{equation}
where we have substituted the value of the parameters $a$ and $c$.
The case $n=2$ gives
\begin{equation}
\label{n=2} q(t) = \sqrt{\frac{E}{\alpha}}\,\sin{\omega (t -
t_0)}\,,
\end{equation}
where the constant $\omega = 1/C_2$ and $t_0=C_1$ remain to be
determined.

\section{Parametric Solutions and Periodicity}
\label{conforme} As we have seen, if $E\neq 0$, the general
solution $t(q)$ cannot be inverted to obtain explicitly $q(t)$. In
such cases, parametric solutions are usually more enlightening.
Some simplicity is achieved when we replace the pair $(t,q)
\rightarrow (\tau, Q)$, where the ``conformal time" $\tau$ and the
new coordinate $Q$ are defined by
\begin{equation}
\label{tau} dt = q(\tau )d\tau \quad,
\end{equation}
and
\begin{equation}
\label{Q} Q=q^{-\frac{n}{2}} \quad.
\end{equation}

Under the above transformations, the expressions for the
Lagrangian and energy E, given by (\ref{lagrangian}) and
(\ref{H}), become
\begin{equation}
L={\frac {2m}{n^{2}}}Q^{-2}Q'^{2} - \alpha Q^{-2} \quad,
\end{equation}

\begin{equation}
E={\frac {2m}{n^{2}}}Q^{-2}Q'^{2} + \alpha Q^{-2}\quad,
\end{equation}
where a prime denotes derivative with respect to $\tau$. As one
may check, the Euler-Lagrange equations are now given by
\begin{equation}
\label{EqMB} QQ'' - Q'^{2} - \frac{\alpha n^{2}}{2m}=0 \quad,
\end{equation}
and inserting $Q'^{2}$ from the energy conservation law, we obtain
\begin{equation}
\label{confoscil} Q'' -\left(\frac{n^2E}{2m}\right)Q= 0 \;.
\end{equation}
This is an interesting result. The above equation describes the
motion of a classical particle under the action of a linear force.
If $E<0$ ($E>0$) the force is of restoring (repulsive) type, while
the motion of a free particle corresponds to $E=0$. The general
solution of (\ref{confoscil}) is
\begin{equation}
\label{confosol} Q = \frac{Q_0}{\sqrt{\epsilon}}
\sin{\sqrt{\epsilon}(\omega \tau + \delta)}\;,
\end{equation}
where \( \epsilon = -2E/m \), \( \omega = n/2 \) is the frequency
of the oscillator (anti-oscillator) and $Q_0$, $\delta$ are
integration constants. Since the energy $E$ may be negative, the
``phase'' $\delta$ may assume complex values. The constant $Q_0$
can be determined using the energy equation written in terms of
the conformal time. It turns out that $Q_0 = \sqrt{-2\alpha /m}$.
Substituting this into (\ref{confosol}), and returning to the
original coordinate $q$, we have
\begin{equation}
\label{confoq} q(\tau)=(-m/2\alpha)^{1/n}\left(
\frac{\sin{\sqrt{\epsilon}\omega \tau}}{\sqrt{\epsilon}}
\right)^{-2/n} \;,
\end{equation}
where we have put $\delta =0$ in (\ref{confosol}).

\section{Relativistic Hamilton-Jacobi Approach}

Let us now consider the relativistic treatment for the homogeneous
potential. For a single-particle the relativistic Lagrangean is
\begin{equation}
\label{L-relat}
 L = - mc^2 \sqrt{ 1 - \frac{\dot{q}^2}{c^2} } - \alpha q^n\;.
\end{equation}
The Hamiltonian function for the particle is given by the general
formula
\begin{equation} \label{H-relat}
{\mathcal{H}}=\dot{q}\frac{\partial L}{\partial\dot{q}}-L =
\sqrt{p^2c^2 + m^2c^4} + \alpha q^n\;,
\end{equation}
where $p$ is the canonical momentum. The Hamilton-Jacobi equation
for $S$ is obtained by replacing in (\ref{H-relat}) $p$ by
$\partial S/\partial q$ and $\mathcal{H}$ by $-(\partial
S/\partial t)$:
\begin{equation} \label{HJ-relat}
c^2\left( \frac{\partial S}{\partial q}\right)^{2} - \left(
\frac{\partial S}{\partial t} + \alpha q^n \right)^2 + m^2c^4=0\;.
\end{equation}
As before, a solution for (\ref{HJ-relat}) can be found separating
the variables in the form
\begin{equation}
S(q,E,t)=W(q,E) - Et\,,
\end{equation}
where the integration constant $E$ is again the total energy. With
the above choice one finds from (\ref{HJ-relat})
\begin{equation}
S=\frac{1}{c}\int\sqrt{(E - \alpha q^n)^2 - m^2c^4}dq - Et\,,
\end{equation}
which should be compared with its classical version (\ref{HPF}).
The second integration constant arises out from the transformation
equation $(\beta=\frac{\partial S}{\partial E})$ and we have
finally
\begin{equation} \label{beta1}
t + \beta =\frac{1}{c}\int\frac{E-\alpha q^n}{\sqrt{(E-\alpha
q^n)^2-m^2c^4}}dq \;.
\end{equation}
Without loss of generality we take $q(t=0)=0$, so that $\beta=0$
in (\ref{beta1}). We next suppose that the potential is a
symmetric function about the origin, that is, we consider only
even values for $n$. Then the motion will be limited between
$V(-A)$ and $V(A)$ where the amplitude is now given by
$A=[(E-mc^2)/\alpha]^{1/n}$, and, from (\ref{beta1}), the period
$T$ is determined by
\begin{equation} \label{T-integral}
T = \frac{4}{c}\int_0^{A} \frac{dq}{\sqrt{ 1 - \left(\frac{m
c^2}{E - \alpha q^n}\right)^2 }}\;.
\end{equation}
Writing the total energy as $E = mc^2(1 + \varepsilon)$, and
considering that the potential energy is small compared to the
rest mass energy $mc^2$, relativistic corrections to second order
in the period $T$ are
\begin{equation} \label{T-aprox}
T = \frac{4}{c}\int_0^A \frac{dq}{\sqrt{2\kappa(A^n - q^n)}}(1 +
\frac{ 3 \kappa}{4}(A^n - q^n))
\end{equation}
where $\kappa=\alpha/mc^2$. Introducing the variable change
$y=(q/A)^n$ the above integral can be rewritten as
\begin{equation}\label{Trel}
T = \frac{4}{c A^{\frac{n}{2} - 1} n \sqrt{ 2 \kappa}} \int_0^1
y^{-(\frac{n-1}{n})} (1 - y)^{-(\frac{1}{2})} dy + \frac{ 3 \kappa
A^{\frac{n +2}{2}} }{c n \sqrt{2\kappa}} \int_0^1
y^{-(\frac{n-1}{n})} (1 - y)^{\frac{1}{2}} dy
\end{equation}
whose values are tabulated in terms of the complete Beta
function\cite{abramowitz}. Substituting for $\kappa$ we write the
period as
\begin{equation} \label{period-relat}
T = 2\sqrt{\frac{2m}{\alpha n^2}}\, A^{1-n/2}B(1/n,1/2) \left[
1+\frac{3}{8}\frac{\alpha A^n }{mc^2}\left(
\frac{2n}{n+2}\right)\right] \;.
\end{equation}
Observe that for $n=2$ (relativistic harmonic oscillator), the
above equation reduces to
\begin{equation}
T=2\pi\sqrt{\frac{m}{2\alpha}}\left[1+ \frac{3}{8}\frac{\alpha A^2
}{mc^2} \right]\,,
\end{equation}
where the term $(3\alpha A^2)/8mc^2$ is the first order
relativistic correction earlier obtained by Synge\cite{synge}
following a different approach (see Appendix).

The general form (\ref{period-relat}) is an interesting one. The
term multiplying the square bracket is just the nonrelativistic
period $T_0$ for the bounded homogeneous potential as given by
(\ref{period}). The remaining term represents the relativistic
correction
\begin{equation} \label{correction}
\frac{\Delta T}{T_0}=\frac{3}{8}\frac{\alpha A^n
}{mc^2}\left(\frac{2n}{n+2}\right)\,,
\end{equation}
and, although it depends of $n$, we see that for $n\gg 1$ it
saturates around $(3\alpha A^n)/4mc^2$ which is twice the
relativistic correction for the harmonic oscillator.

In conclusion, we have shown that the classical version of
homogeneous potential problem can completely be solved by the HJ
approach. In the relativistic case we not succeed in obtaining an
analytic solution, but the earlier Lagrangian Synge's treatment
for the harmonic oscillator is readily generalized using the HJ
transformation. Finally, we stress that although important from
their own right, such problems may also have interest for
describing some excited states appearing in quantum chromodynamics
which have been usually described by the relativistic
oscillator\cite{DB83}.

\vskip 0.5cm

{\raggedright {\bf ACKNOWLEDGMENT} }

This work was partially supported by CNPq and CAPES (Brazilian
Research Agencies).

\section{Appendix: Remarks on Synge's Integral}

The relativistic harmonic oscillator ($n=2$) was long ago
discussed by Synge \cite{synge}. In this appendix, we show how the
exact Synge's integral form for the period of the motion can be
generalized for a power-law index.

For an arbitrary $n$, the period is given by (see
(\ref{T-integral}))

\begin{equation} \label{T1}
T_n = \frac{4}{c}\int_0^{A} \frac{dq}{\sqrt{ 1 - \left(\frac{m
c^2}{E - \alpha q^n}\right)^2 }}\;.
\end{equation}

Now, by writing $E = mc^{2} + \alpha A^{n}$ one finds

\begin{equation}
T_n = 2\sqrt{\frac{2m}{\alpha A^n}} \int^{A}_{0} \frac{[1 +
2\chi^2(1 - (q/A)^n )]\,dq}{\sqrt{1 - (q/A)^n + \chi^2 (1 -
(q/A)^n )^2}}\,,
\end{equation}
where $\chi^2 = \alpha A^n/(2mc^2)$.

Introducing the variable $\varphi$ by $q = A\sin^{2/n} \varphi$,
the above integral becomes

\begin{equation}\label{TG}
T_n = 4\sqrt{\frac{2m}{\alpha n^2}}\,A^{1-n/2}
\int^{\frac{\pi}{2}}_{0} \frac{\sin^{2/n - 1}\varphi\,(1 +
2\chi^2\cos^2 \varphi)\,d\varphi}{\sqrt{1 + \chi^2
\cos^2\varphi}}\,
\end{equation}
and, for $n=2$ we have
\begin{equation}
T_2 = 4\sqrt{\frac{m}{2\alpha}} \int^{\frac{\pi}{2}}_{0} \frac{(1
+ 2\chi^2\cos^2 \varphi)d\varphi}{\sqrt{1 + \chi^2
\cos^2\varphi}}\;,
\end{equation}
which is the Synge's integral with a slight different notation
(our $\alpha = k^{2}/2$, and $A=a$ in Synge's work\cite{synge}).
Finally, by expanding (\ref{TG}) in powers of $\chi$, the
relativistic correction for the period as given by
(\ref{period-relat}) is recovered. For n=2 it reduces to Synge's
expression.


\begin{thebibliography}{99}

\bibitem{goldstein}
H. Goldstein, {\em Classical Mechanics\/} (Addison-Wesley Publ.
Co., Reading, Mass. 1980), pp. 438-498.

\bibitem{Holland} P. R. Holland, {\em The Quantum Theory of
Motion} (Cambridge UP, Cambridge, 1995), pp. 27-65.

\bibitem{Carter} B. Carter, Phys. Lett. {\bf 26 A},
399-400 (1968); Phys. Rev. {\bf 174}, 1559-1571 (1968).

\bibitem{LL}
L. D. Landau and E. M. Lifshitz, The Classical Theory of Fields,
(Pergamon Press, New York, 1975), pp. 328-330.

\bibitem{FVL} W. H. C. Freire, V. B. Bezerra, and J. A. S Lima,
{\bf GRG 33}, 1407-1414 (2001); R. C. Santos, J. A. S. Lima, and
V. B. Bezerra, {\bf GRG 33}, 1969-1977 (2002).

\bibitem{Valle96} G. del-Valle, I. Campos, and J. L. Jim\'enez, Eur. J.
Phys. {\bf 17}, 253- (1996); I. Campos, G. del-Valle, J. L.
Jim\'enez, Eur. J. Phys. {\bf 24}, 469-479 (2003).

\bibitem{pars}
L. A. Pars, A Treatise on Analytical Dynamics, John Wiley \& Sons,
New York (1965).

\bibitem{synge} J. L. Synge, Classical Dynamics, in Encyclopedia of
Physics, Vol. III, 210-214 (1960).

\bibitem{abramowitz}
M. Abramowitz and I. A. Stegun, Handbook of Mathematical Functions
(Dover, New York, 1970), pp. 255.

\bibitem{limajas}
J. A. S. de Lima, J. A. M. Moreira, and J. Santos, {\bf GRG 30},
425-434 (1998).

\bibitem{limajas2} M. J. D. Assad and J. A. S. Lima, {\bf GRG 20}, 427
(1988).

\bibitem{seaborn}
J. B. Seaborn, {\em Hypergeometric Functions and Their
Applications\/} (Springer-Verlag, New York, 1991), pp.

\bibitem{DB83} C. Dullemond and E. van Beveren, Phys. Rev. {\bf D 28},
1028-1032 (1983).

\end{thebibliography}
\end{document}